\documentclass[11pt]{article}
\usepackage{amstex}
\title{\Large{\bf The Mathematical Theory of Molecular Motor Movement and 
Chemomechanical Energy Transduction} \\[8pt]
\normalsize{Hong Qian} \\[8pt] 
Departments of Applied Mathematics and Bioengineering \\
University of Washington, Seattle, WA 98195, USA}
\begin{document}

\maketitle

\begin{abstract}
The mathematical formulation of the model for molecular
movement of single motor proteins driven by cyclic 
biochemical reactions in an aqueous environment leads to a 
drifted Brownian motion characterized by coupled diffusion 
equations.  In this article, we introduce the basic notion 
for the continuous model and review some asymptotic solutions 
for the problem. (For the lattice model see \cite{FK,Q3}.)
Stochastic, nonequilibrium thermodynamic 
interpretations of the mathematical equations and their 
solutions are presented.  Some relevant mathematics,
mainly in the field of stochastic processes, are 
discussed.
\end{abstract}

\def\calA{$\mathcal{A}$}
\def\calM{$\mathcal{M}$}
\def\calS{$\mathcal{S}$}
\def\AmSTeX{{\calA\kern-.1667em%
  \lower.5ex\hbox{\calM}\kern-.125em\calS}-\TeX}
\def\rvX{{\bf X}}
\def\vecxi{\hbox{\boldmath $\xi$}}
\def\wt{\widetilde}
\def\opL{{\mathfrak L}}
\section{Introduction}
\noindent

One of the fascinating aspects of protein molecules in the biological
world is their ability to perform various, almost ``magic-like'' 
tasks \cite{Fer}.  A particular class of proteins known as molecular motors 
can move linearly along its designated  track against an external force 
by utilizing the biochemical energy source, adenosine triphosphate (ATP).
In this manner, the motor proteins act as miniature engines converting 
chemical energy to mechanical work.  Movement of single protein molecules 
inside a cell, however, has to experience thermal agitation from the 
aqueous environment in the cytosol.  The movement is therefore a Brownian 
motion with drift (convective diffusion) \cite{Ber}.  

	Such movement provides the molecular basis for muscle 
contraction and various cellular transport processes \cite{Ho,Hu}.  
Motor protein kinesin is known to carry out intra-cellular 
vesicle transport along microtubules.  Various polymerases 
are moving along their corresponding templates.  All these processes
are essential to a living cell.  In a muscle cell, 
the motor protein is called myosin, and its designated track is called 
an actin filament.  The actin filament has a periodic structure of 
$\sim$36nm.  Therefore, without loss of generality, we assume that a
myosin molecule moves in a force field with periodic potential energy 
function $U(x)$: $U(x+L) = U(x)$ where $L$ is 36nm for actin.  

	Treating the center of mass of the motor protein as a Brownian
motion with the presence of a periodic energy potential, its movement 
can be modeled by the Smoluchowski equation \cite{Wa}
\begin{equation}
  \frac{\partial P(x,t)}{\partial t} = -\frac{\partial}{\partial x} J(x,t)
	= D\frac{\partial^2 P(x,t)}{\partial x^2} 
     - \frac{\partial}{\partial x} \left( \frac{F(x)}{\beta} P(x,t)\right),
\label{ahe0}
\end{equation}	
where $D$ and $\beta$ are, respectively, the diffusion and frictional
coefficients. $F(x) = -dU(x)/dx$ is the force of the potential
$U$, representing the molecular interaction between the motor protein and 
its track.  $P(x,t)$ is the probability density function of the motor protein 
at position $x$ for time $t$.  The first equality in Eqn. (\ref{ahe0}) is 
a continuity equation in which $J$ is a probability flux.  The first 
term on the right-hand-side is associated with the diffusion flux 
according to Fick's law. The second term is due to the convection 
associated with an overdamped Newtonian motion: $\beta \dot{x}=F(x)$.   
In fact, Eqn. (\ref{ahe0}) is mathematically equivalent to an overdamped
Newtonian motion with a white random force $f(t)$ representing the 
incessant collisions between the motor protein and the water
molecules: $\beta \dot{x}=F(x)+f(t)$ \cite{Wa}. 

	The probability density $P(x,t)$, as the solution to Eqn.
(\ref{ahe0}), gives the mean position of the motor
\begin{equation}
   \langle x \rangle(t) = \int_{-\infty}^{\infty} x P(x,t) dx.
\end{equation}
Moreover, its velocity is related to the flux $J(x,t)$:
\begin{eqnarray}
 \frac{d}{dt} \langle x \rangle(t) 
    &=& \int_{-\infty}^{\infty} x \frac{\partial P(x,t)}{\partial t} dx 
 						\nonumber\\
    &=& -\int_{-\infty}^{\infty} x \frac{\partial J(x,t)}{\partial x} dx
       						\nonumber\\
    &=& \int_{-\infty}^{\infty} J(x,t)dx.	\label{vel}
\end{eqnarray} 
When the motion of a motor protein becomes steady and if we are
only interested in the mean velocity, we need only consider 
the steady-state solution for $x \in [0,L]$ with periodic boundary 
conditions.  With this setting, the steady-state velocity of the 
motor protein movement is $v=LJ$.  Therefore studies on
the steady-state movement focus on the flux $J$.  To fix our terminology,
we will refer to a stationary solution as a steady-state, but a 
stationary solution with zero flux as an equilibrium. 

	A little mathematical analysis immediately shows that with
$U(0)=U(L)$, i.e., $\int_0^L F(x)dx = 0$, the stationary solution of 
(\ref{ahe0}) allows only zero flux ($J=0$). {\it Therefore, in a 
one-dimensional periodic structure, there is no driving force to 
bias an inert Brownian particle to move in either direction.} 
The simplest model given by Eqn. (\ref{ahe0}) fails to capture
the essence of motor protein movement.

	The driving force for a motor protein comes from a very 
important biochemical reaction, occurring inside the protein, 
called ATP hydrolysis:
\vskip 0.15cm
\centerline{ATP + H$_2$O 
$\overset{f}{\underset{g}{\rightleftharpoons}}$ 
ADP + Pi} 
\vskip 0.15cm \noindent
where H$_2$O is water, ADP is adenosine diphosphate, and Pi is phosphate.  
Chemical reactions like this are well characterized by a two-state Markov
process (or more generally, $m$ discrete states)
\begin{eqnarray}
     \frac{dP_{ATP}}{dt} &=& -f P_{ATP} + g P_{ADP}  \nonumber\\
     \frac{dP_{ADP}}{dt} &=& f P_{ATP} - g P_{ADP}   \label{ATPh} 
\label{ATP}
\end{eqnarray}
where non-negative $f$ and $g$ are rate constants for the reaction 
\cite{Mc}.  The deeply insightful work of Sir Andrew Huxley in 1957 was to
introduce internal conformational states to the Brownian particle 
and to couple the (bio)chemical reaction in (\ref{ATP}) with the 
motor protein movement in (\ref{ahe0}) \cite{Hu}.  This leads to the 
following equation \cite{AB2,PEO,Q2,JAP}, known as a coupled 
diffusion system in mathematics \cite{QZ}, for the 
movement of a Brownian particle with internal structures and
dynamics:
\small
\begin{eqnarray}
   \frac{\partial P(x,-)}{\partial t} &=& 
		D_-\frac{\partial^2 P(x,-)}{\partial x^2} 
     +\frac{1}{\beta_-}\frac{\partial}{\partial x}
       \left( \frac{dU_-}{dx}P(x,-)\right)
	      - f(x)P(x,-) + g(x)P(x,+)  \nonumber  \\
   \frac{\partial P(x,+)}{\partial t} &=& 
		D_+\frac{\partial^2 P(x,+)}{\partial x^2} 
     + \frac{1}{\beta_+}\frac{\partial}{\partial x}
       \left( \frac{dU_+}{dx}P(x,+)\right)
                 + f(x)P(x,-) - g(x)P(x,+),         \nonumber
\end{eqnarray}  
\normalsize  
\begin{equation}
\label{ahe1}
\end{equation}
where $f(x)$ and $g(x)$ are non-negative periodic functions. 
In terms of this augmented Huxley model, the motor protein can be 
either attached to (+) or detached from ($-$) the filament
with respective interaction energy functions $U_+(x)$ and $U_-(x)$
(In the original Huxle model, $U_-(x)=0$).  $D_{\pm}$ ($\beta_{\pm}$)
are the diffusion (frictional) coefficients of the motor protein in 
the attached and detached states, respectively.  The attach-detach 
transition is coupled to the ATP hydrolysis.  Therefore, in (\ref{ahe1})
the biochemical reaction is coupled to the movement of the motor
protein.  More importantly, note that in either the attached or detached 
state, there is no bias for the motor protein movement!  {\it However, 
the attach-detach transition driven by the ATP hydrolysis leads to a 
biased motion of the motor protein (J$\neq$0).  The chemical energy 
in ATP is converted to the mechanical motion of the motor protein.}
This we shall show.

	At this point, it is fascinating to read the now classic
work of Huxley on the theory of muscle contraction, which was written
decades before the discovery of the motor protein molecule in its
individual form.  The Huxley model works as follows (p. 281 of
\cite{Hu}).  Initially, the myosin (M) and actin filament (A) are 
detached; M oscillates (fluctuates) back and forth about its equilibrium
position $O$ as a result of thermal agitation (with diffusion
coefficient $D_-$ for the Brownian motion).  If A happens to be
within the range of positions where $f$, the rate of association,
is not zero, there is a chance that combination will take place
(this event happens with a probability characterized by the Markov 
rate process); when this has happened the tension in the elastic element
($F$, i.e., the molecular interaction between M and A) will
be exerted on the actin thread by M (and, conversely, the actin
will exert the same force with opposite sign on M).  As one can see, 
Eqn. (\ref{ahe1}) is the mathematical formulation of this model described
in words.  Readers who check Huxley's paper will find several 
differences between his original equation and (\ref{ahe1}).  
These differences arise because we have formulated a 
{\it microscopic} model for single motor proteins while a model
for muscle contraction has to deal with a large number of myosin
molecules.  It can be shown that when stringing many 
motor proteins into a rigid chain, Huxley's original equation 
can be derived from Eqn. (\ref{ahe1}) \cite{JAP}.

	Since the work of Huxley, there have been many investigations 
following and expanding the basic notion of the Huxley equation.  Most 
notably are the work of Hill \cite{Hi1,Hi2} who provided the Huxley 
equation with a sound thermodynamic basis, and the work of 
Astumian and Bier, and Peskin et al. who arrived at (\ref{ahe1}) from 
the Langevin dynamics (stochastic differential equation) 
point of view \cite{AB1,PEO}.  There is also a body of literature 
on Brownian ratchet, whose basic movement is characterized by equations 
identical to (\ref{ahe1}) \cite{As,Ch,DHR,POO,ZC}.\footnote{It is
important here to qualify the term ``Brownian ratchet'' which
does not involve any temperature gradience \cite{PEO}.  It is 
an isothermal device in which useful work is derived from 
non-equilibrium fluctuations.  It violates the detailed balance 
due to active pumping \cite{Q1,Q2}.}  In a recent paper, 
Bentil \cite{Ben} applied the Huxley model in conjunction with 
Langevin dynamics to simulate single myosin experiments.  
Qian \cite{Q2,Q3,Q4} established a relationship between coupled 
diffusion like (\ref{ahe1}) and the circulation of a Markov 
process and its entropy production \cite{Kal,Q5}. 

	Eqn. (\ref{ahe1}) is certainly an oversimplified model for
any realistic biological system.  However, it captures the essence
of a theory which unifies microscopic motor protein movement and 
macroscopic muscle contraction; thereby it provides a concrete model
for chemomechanical energy transduction in living organism.  Hence
it deserves further detailed investigation as a topic in biophysics
and physical chemistry. The mathematical treatment of (\ref{ahe1}) 
has been mainly in terms of differential equations.  However, the
nature of Brownian motion of the motor protein also calls for a
treatment of (\ref{ahe1}) in terms of stochastic processes.
We will review some of the pertinent mathematics in Section 4.

\section{Mathematical Analyses of Several Limiting Cases of 
Augmented Huxley Equation}
\noindent

	While the augmented Huxley Eqn. (\ref{ahe1}) is difficult 
to solve in general due to the non-local (non-equilibrium) nature 
of the steady-state, particular limiting cases can be analyzed to 
gain insights into the theoretical model.  In this section, we 
present some known and also some new results.

\subsection{Limit of rapid biochemical cycling} 

	One particularly interesting limiting case is when the 
biochemical reactions are rapid with respect to the diffusion.
Analysis of this limiting case clearly demonstrates how the
internal biochemical reaction can give rise to a unidirectional
motor protein movement.  Hence it demonstrates the validity 
of mathematical models for motor proteins in terms of coupled 
diffusion equations.

	Consider Eqn. (\ref{ahe1}), rapid biochemical reaction 
means we have conditional probabilities:\footnote
{Readers who are familiar with the method 
of singular perturbation will identify this problem.  Since we 
are seeking a nontrivial solution for a homogeneous equation, 
the solution is not unique.  The functions $F_{\pm}$ has at least 
two zeros at which boundary layers might be expected.  For more detail
on this type of equations, see \cite{OW}.}
\[     P(-|x) = \frac{g(x)}{f(x)+g(x)}, \hspace{0.5cm}
       P(+|x) = \frac{f(x)}{f(x)+g(x)}                 \]
and thus  $P(x) = P(x,-)+P(x,+)$ satisfies
\begin{equation}
  \frac{\partial P(x)}{\partial t} = 
  \frac{\partial^2}{\partial x^2}(\overline{D}(x)P(x,t))
           - \frac{\partial}{\partial x} 
                 (\overline{F}(x) P(x)) 
\label{meanF}
\end{equation}	
in which
\[  \overline{D}(x) = \frac{f(x)D_++g(x)D_-}{f(x)+g(x)},
    \hspace{0.5cm}
    \overline{F}(x) = \frac{f(x)F_+(x)+g(x)F_-(x)}{f(x)+g(x)}.    \]
Note that Eqn (\ref{meanF}) is similar to Eqn. (\ref{ahe0}) but with
one crucial difference: The mean force function now satisfies
$\int_0^L \overline{F}(x)dx \neq 0$ even though both $F_{\pm}(x)$
satisfy $\int_0^L F_{\pm}(x)dx = 0$.
{\it The potential of the average of forces of periodic L potentials
is in general not periodic.}  This indicates that 
{\it the biochemical reaction (\ref{ATPh}) provides a drift for the 
motor protein movement in a periodic system}.  Finally, the transport 
flux can be obtained by solving (\ref{meanF})
\begin{equation}
  \Phi = \frac{1-e^{u(L)}}
     {\int_0^L e^{-u(x)}dx\int_0^Le^{u(x)}dx -
     (1-e^{u(L)})\int_0^L e^{-u(x)}dx\int_0^x e^{u(x')}dx'}.
\end{equation}
where $u(x)=-\int_0^x \overline{F}(x)dx$.  Therefore, if
$\int_0^L \overline{F}(x)dx = -u(L)>0$, $\Phi >0$.

\subsection{Limit of rapid diffusion} 

	Another limiting case which has been nicely analyzed by 
Peskin et al \cite{PEO} is when the diffusion is very 
rapid in comparison to the Markovian transition (Brownian
ratchet): there are rapid equilibria for $P(x,-)$ and $P(x,+)$.

	The mathematical problem is framed as follows.  Let's 
consider the stationary coupled diffusion:
\begin{eqnarray}
		\frac{d^2 P(x,-)}{dx^2} 
     + \frac{d}{dx}
       \left(\frac{dU_-(x)}{dx}P(x,-)\right)
     -\epsilon (f(x)P(x,-) - g(x)P(x,+)) &=& 0 \nonumber\\
		\frac{d^2 P(x,+)}{dx^2} 
     + \frac{d}{dx}
       \left(\frac{dU_+(x)}{dx}P(x,+)\right)
     +\epsilon (f(x)P(x,-) - g(x)P(x,+)) &=& 0, \nonumber\\
\end{eqnarray}
where $f(x)$, $g(x)$, $U_{\pm}(x)$ are periodic functions.
When the regular perturbation parameter $\epsilon=0$, this
system of uncoupled diffusion has zero transport flux $\Phi$.  
For small $\epsilon$, we therefore have $\Phi = \epsilon\phi+...$ 
and the asymptotics can be obtained by the method of regular 
perturbation \cite{BO}.  At this point it is important to
notice the other flux, the circular flux $\Pi$ which
moves the motor protein forward in the (+)-state but moves 
it backward in the ($-$)-state.  Therefore $\Pi$ does not contribute 
to the net transport but only generates heat.  This type of flux is
known as futile cycle in muscle biochemistry \cite{Q1}.

	Perturbation calculations show that \cite{PEO}
\begin{equation}
  \Pi = \epsilon \frac{\int_0^L g(x)\mu_+(x)dx\int_0^L f(x)\mu_-(x)dx}
             {\int_0^L g(x)\mu_+(x)dx+\int_0^L f(x)\mu_-(x)dx} +
              O(\epsilon^2),
\end{equation}
and
\begin{equation}
  \Phi = \epsilon \int_0^L dx [\nu_+(x)-\nu_-(x)]
                  \int_0^x dx'[f(x')\mu_-(x')-g(x')\mu_+(x')]
       + O(\epsilon^2),
\end{equation}
where
\[ \mu_- = \frac{e^{-U_-(x)}}{\int_0^L e^{-U_-(x)} dx},\hspace{0.3cm}
   \mu_+ = \frac{e^{-U_+(x)}}{\int_0^L e^{-U_+(x)} dx},  \]
\[ \nu_- = \frac{e^{U_-(x)}}{\int_0^L e^{U_-(x)} dx}, \hspace{0.3cm}
   \nu_+ = \frac{e^{U_+(x)}}{\int_0^L e^{U_+(x)} dx}.   \]

	It is interesting to note that if
$f(x)\mu_-(x)-g(x)\mu_+(x)=0$ $\forall$ $x$, then the system is 
reversible and the steady-state is in fact an (thermal) equilibrium.
In mathematical term, the Markov process is symmetric \cite{Si}.  
In applied mathematics, the symmetricity leads to the 
Grasman-Matkowsky variational method \cite{Ke}.  More recent
progress can be found in \cite{KC}.

\subsection{Limit of the original Huxley model}

	In the original Huxley model, the interaction between
the track and the motor in the detached state is assumed to be 
zero.  Hence $U_-(x)=0$ in Eqn. (\ref{ahe1}).  Furthermore, it is 
also generally accepted that $D_+ << D_-$, i.e., the Brownian 
motion of the motor protein in the attached state is negligible.  
The solution of (\ref{ahe1}) when $D_+ \rightarrow 0$ is a 
problem of singular perturbations \cite{Ke,Mu}.  Note that because 
of the periodic $U(x)$, $F(x)$ has zeros on $[0,L]$.  Thus the 
singular perturbation problem has at least two linear turning 
points \cite{Kam,Om}.  The reduced equation when $D_+=0$ is:  
\begin{eqnarray}
       \frac{d^2 P(x,-)}{dx^2} 
	      - f(x)P(x,-) + g(x)P(x,+) &=& 0 		\nonumber\\
     - \frac{d}{dx}
       \left( F(x) P(x,+)\right)
                 + f(x)P(x,-) - g(x)P(x,+)  &=& 0,        \label{ahe2}
\end{eqnarray}    
in which we have set $D_-=\beta_+=1$ for simplicity. 
We are particularly interested in finding a condition for the 
existence of a solution corresponding to unidirectional motion.
In the steady-state, the {\it total transport} flux of the system is 
a constant:\footnote{We use $\Phi$ to denote the transport flux 
$J_+ +J_-$.  As will be shown below there is another type of flux,
circular and non-transport $\Pi = J_+-J_-$, in these systems \cite{Q2}.}
\begin{equation}
    F(x)P(x,+) - \frac{dP(x,-)}{dx} \equiv \Phi
\label{fluxJ}
\end{equation}
Using Eqn. (\ref{fluxJ}) and eliminating $P(x,+)$ from 
Eq. (\ref{ahe2}) we then have 
\begin{equation}
        F(x) \frac{d^2 P(x,-)}{dx^2} 
     + g(x) \frac{dP(x,-)}{dx} 
     - F(x)f(x)P(x,-) =  - \Phi g(x),   
\label{ss}
\end{equation}
where the inhomogeneous term $\Phi$ on the rhs is to be determined 
by the normalization condition $\int_0^L [P(x,+)+P(x,-)]dx=1$.  
The boundary conditions for Eq. (\ref{ss}) again are periodic.  

	Eqn. (\ref{ss}) has singular points at the zeros of $F(x)$.
A simple local analysis shows that for nonzero $\Phi$ and a physically
meaningful $P(x,-) \ge 0$,  the solution to Eqn. (\ref{ss}) 
has to be nonanalytic at these singularities. This nonanalytic 
behavior, however, is expected to be obviated in an asymptotic
study of the full equation (\ref{ahe1}) with small $D_+$.

\section{Entropy Production in Nonequilibrium Steady-State} 
\noindent

	We now give a brief discussion of the nonequilibrium
thermodynamics in terms of Eqn. (\ref{ahe1}). 
Hill \cite{Hi1,Hi2} has given an extensive account 
of this subject.  We only discuss some recent developments in 
connection with the notion of entropy production \cite{On}.  
The concept of entropy production rate (e.p.r.) can be easily
introduced, mathematically, in terms of Eqn. (\ref{ahe0}). 
The validity of this novel thermodynamics of nonequilibrium
steady-state (NESS), however, remains to be experimentally tested.  
For more discussion see \cite{Q2,Q3,QW,QQG}.  

	Associated with (\ref{ahe0}) is a functional $A[P(x)]$ 
called the Helmholtz free energy in thermal physics \cite{Reic}, 
which in units $k_BT$ ($T$ is temperature and $k_B$ is the 
Boltzmann constant) is defined as:
\begin{equation}
    A[P(x,t)] = \int_0^L U(x)P(x,t)dx + \int_0^L P(x,t)\log P(x,t) dx.
\end{equation}
When $P(x,t)$ changes with time according to Eqn. (\ref{ahe0}),
the production rate of total entropy is the rate of decrease 
in $A$ of the system,\footnote{From thermodynamics stand point, a
macromolecule is an isothermal system in contact with a thermal 
environment (i.e., aqueous solution) with temperature $T$.  
There is energy (chemical and heat), but no material, exchange 
between the system and its environment (clamped ATP and ADP
concentration in the aqueous solution, and heat bath).  The 
system and its environment as a whole is an isolated system 
with a constant total energy; and this is approximately hold also 
for a sufficiently large heat bath.  In mathematical terms, we have 
$dA/dt = dE/dt - TdS/dt$ where $E$ is the internal energy of the 
system and $S$ is the entropy of the system.  With respect to 
the system and the environment together as a whole $dA$ = 
$(dE_{tot}-dE_{env})-T(dS_{tot}-dS_{env})$ =  
$-TdS_{tot}+(TdS_{env}-dE_{env})$ $\approx$ $-TdS_{tot}$.  
For an isolated system (microcanonical ensemble),
$\partial E_{env}/\partial S_{env} = T$ \cite{Reic}.} 
which can be computed:
\begin{eqnarray}
     e.p.r. =
      -\frac{dA}{dt} &=& -\int_0^L U(x)\frac{dP(x)}{dt}dx - 
                 \int_0^L \frac{dP(x)}{dt} \log P(x) dx  \nonumber\\
   &=& \int_0^L (U(x)+\log P(x))\frac{d}{dx}J(x)dx     \nonumber\\
   &=& \int_0^L \left((F(x)-\frac{d}{dx}\log P(x)\right)J(x)dx
							  \nonumber\\
   &=& \frac{1}{D} \int_0^L J^2(x)P^{-1}(x) dx  \ge 0. 	  \label{ep} 
\end{eqnarray}
The last step used the definition for $J$ given in Eq. (\ref{ahe0}).
This is the second-law of thermodynamics in terms of 
Smoluchowski equation, which is the nonequilibrium counterpart of 
a canonical ensemble in statistical mechanics.  Eqn. (\ref{ep})
can be generalized to calculating the entropy production rate
(e.p.r.), as well as the heat dissipation rate, in a 
NESS in which $S_{tot}$ continues to
increase \cite{Q7}.  To see this, we note the entropy of the 
system is defined as:
\begin{equation}
    S[P(x,t)] = -\int_0^L P(x,t)\log P(x,t) dx.
\end{equation}
Therefore in NESS, 
\begin{equation}
    \dot{S} = -\int_0^L FJ dx + e.p.r = 0.
\end{equation}
where $F=-dU/dx$, and the first term on the right-hand-side
is the heat dissipation rate (h.d.r.).  Therefore, in a
NESS, the e.p.r. is equal to the h.d.r.

	Recent work in mathematical physics on entropy 
production in nonequilibrium systems \cite{GC} also 
focus on appropriately setting up the non-equilibrium 
steady-state with an external force and a thermostat simultaneously
acting on a Hamiltonian system.  The force supplies energy while 
the thermostat removes heat in order to keep the system in a 
steady-state with bounded energy.  This leads to a {\it random 
dynamical system} in which entropy production is cogently defined \cite{Ru}. 
The Smoluchowski approach we adopt has a quite similar setting: 
a driving force due to chemical reaction (rather than mechanical force) 
and an implicit thermostat: The Smoluchowski equation is a consequence 
of an overdamped Newtonian system with Maxwellian distribution for
the velocity of the particles \cite{Wa}.  In fact, the diffusion 
coefficient $D$ and frictional coefficient $\beta$ in (\ref{ahe0}) 
define the temperature of the thermostat: $T=\beta D/k_B$.
The interesting mathematical questions are when these 
random dynamical systems become diffusion processes and whether the 
entropy production proposed in these studies is equivalent to 
Eqn. (\ref{ep}) for diffusion processes \cite{QW}.  
The recent work by Lebowitz and Spohn \cite{LS} has provided some 
insights on this problem \cite{Q5}.
In a different approach to weak random perturbation of Hamiltonian 
systems in a plane, Freidlin and Wentzell map the system 
to a diffusion process on a graph, which consists of vertices 
corresponding to the stationary states and edges corresponding
to energy basins \cite{FW}.  However, this approach remains to
be generalized to higher dimensional Hamiltonian 
systems, and its relationship to the (Kramers') 
transition-state rate theory in theoretical chemistry \cite{HTB} 
also remains to be elucidated. 

	It is interesting to note that in a steady-state, the 
flux $J$ is a constant and all the entropy produced will 
become the dissipated heat.  Hence 
e.p.r.$=(J^2/D)\int_0^L P^{-1}(x)dx$ $\ge$ $J^2L^2/D$ which 
equals $\beta J^2L^2$ due to the Einsteine relation in $k_BT$ units
$D\beta =1$.  The $\beta J^2L^2$ term is the energy dissipation 
due to a deterministic motion with velocity 
$v=LJ$ and frictional coefficient $\beta$ in continuous medium. 
The inequality indicates the additional dissipation due to 
random motion.   It also indicates that when $P(x) =$ 
constant, the e.p.r is at its minimum, i.e., the 
chemomechanical energy transduction is at its maximal
efficiency. 

	To generalize the concept of entropy production
to Eqn. (\ref{ahe1}) is mathematically straightforward.  
This yields a novel thermodynamic theory for NESS, which is 
particularly relevant to motor proteins. The importance of 
the theory is that it relates e.p.r. to the heat 
production of a working motor, which is a quantity that 
can be experimentally measured.  With some simple algebra, 
we have
\begin{equation}
  h.d.r. = \int_0^L \left[F_+(x)J_+(x)+F_-(x)J_-(x)
	 + j(x)\log\left(\frac{f(x)}{g(x)}\right) \right]dx
\label{hdr}
\end{equation}
\begin{equation}
     e.p.r. = \int_0^L \left[J_+^2(x)P(x,+)+J_-^2(x)P(x,-)
	    + j(x)\log\left(\frac{f(x)P(x,-)}{g(x)P(x,+)}\right) 
		\right]dx
\label{epr2}
\end{equation}
in which 
\[    J_{\pm}(x) = -\frac{dP(x,\pm)}{dx}+F_{\pm}(x)P(x,\pm),   \]
and 
\[    j(x) = f(x)P(x,-)-g(x)P(x,+).  \]
It is obvious that e.p.r. is non-negative and equal to zero
if and only if when detailed balance is hold \cite{JAP,Q4}. 
In NESS without external load, h.d.r. = e.p.r.

	If there is an external load $F_{ext}$, then Eqn. (\ref{hdr})
can be further broken down into 
\begin{equation}
     k_BT \ln\left(\frac{[ATP]}{[ADP][Pi]}e^{\Delta G^0_{ATP}}\right)
          \int_0^L j(x)dx - F_{ext}(J_++J_-)L.
\end{equation}
Hence, The free energy from ATP hydrolysis is equal to exactly 
the sum of the work done against the external load, $F_{ext}v$,
and positive heat dissipation, Eqn. (\ref{epr2}).

\section{Some Relevant Mathematics on Coupled Diffusion}  
\noindent

	While $P(x,t)$ in Eqn.  (\ref{ahe0}) characterizes a 
stochastic process $\rvX_t$ in terms of probability 
density at each time $t$, $P(x,t)dx$ = 
$Prob\{x\le\rvX_t\le x+dx\}$, there is an alternative view of a 
stochastic process in terms of its {\it trajectories}.  In this
approach, all possible trajectories $\{\rvX_t|t\ge 0\}$ form a
function space $\Omega$ and a probability density (a measure)
is defined.  This naturally leads to the notion of ``propagator''
(a semi-group) which is formally defined as:
\begin{equation}
         P(x,t) = P(x,0)e^{\opL t}  
\end{equation} 
where the exponential operator $e^{\opL t}$ acts on the
distribution $P(x,0)$ as a ``row vector''.  The operator $\opL$ 
satisfies the backward Kolmogorov equation: 
\begin{equation}
     \frac{\partial P(x,t)}{\partial t} = P(x,t)\opL,
\end{equation}
or its conjugate $\opL^*$ satisfies the 
forward Kolmogorov (or Fokker-Planck) equation: 
\begin{equation}
     \frac{\partial P(x,t)}{\partial t} = \opL^* P(x,t).
\label{difeqn}
\end{equation}
For symmetric operator $\opL^* = \opL$.
The symbolic relationship between the operator $\opL$ and 
the propagator $e^{\opL t}$ has been made rigorous in terms of
linear operators in a Banach space and is known now as Hille-Yosida 
theorem.  Hence the modern theory of Brownian motion has brought 
several mathematical disciplines to bear \cite{Fel,KS}: partial 
differential equations, linear operators on functional spaces, 
and harmonic analysis.  

\subsection{Feynman-Kac formula}  

	A major result in this area is a relationship 
between the solution of a boundary value 
problem (BVP) and the mean exit time (first passage time) of a diffusion 
process.  This relation also has the potential for devising numerical 
methods for solving BVP.  Let's now consider a Brownian motion in a domain 
$D$ with the diffusion equation (\ref{difeqn})
and Dirichlet boundary conditions on $\partial D$. 
Now differentiate Eq. (\ref{difeqn}) with respect to $t$, and 
then multiply a $t$ and integrate over $t \in (0,\infty)$, we
have
\begin{equation}
       \int_0^{\infty} t \frac{\partial^2 P(x,t)}{\partial t^2} dt
      = \opL^* \int_0^{\infty} t 
                  \frac{\partial P(x,t)}{\partial t} dt.
\label{difeqn2}
\end{equation}
If one interprets $P(x,t|x_0)$ as the probability of 
the Brownian particle at $x$ at time $t$ starting at $x_0$ when 
$t=0$, then
\[   u(x_0) = -\int_D dx \int_0^{\infty} t 
                   \frac{\partial P(x,t|x_0)}{\partial t} dt    \] 
is the mean time of the particle started at $x_0$ to exit $D$.
The left-hand-side of (\ref{difeqn2}) can be simplified into 
\[    \left[ t\frac{\partial P(x,t)}{\partial t} - P(x,t) 
      \right]_0^{\infty} = \delta(x-x_0).    \]
If now we multiply a function $\phi(x)$, which satisfies 
$\opL\phi(x)=\psi(x)$ with (\ref{difeqn2}) and then integrate 
over $x \in D$, we have
\begin{eqnarray*}
    \phi(x_0) &=& \int_D dx \phi(x) \opL^* \int_0^{\infty} t 
                  \frac{\partial P(x,t)}{\partial t} dt       \\
              &=& \int_D dx (\opL\phi(x)) \int_0^{\infty} t 
                  \frac{\partial P(x,t)}{\partial t} dt       \\
              &=& \int_D \psi(x) dx \int_0^{\infty} t 
                  \frac{\partial P(x,t)}{\partial t} dt       \\
              &=& -E^{x_0} \left[\int_0^{\tau_D} \psi(\rvX_s)ds\right]
\end{eqnarray*}                
This shows that the rhs, in which $\rvX_t$ is the probabilistic Brownian 
motion and $E^{x_0}[\cdot]$ is the average along the paths of $\rvX_t$ started 
at $x_0$, satisfies the inhomogeneous ODE
\begin{equation}
           \opL u(x) = -\psi(x).     
\end{equation}
This is the well-known Feynman-Kac formula \cite{Ok}.  When $\psi(x)\equiv 1$, 
$u(x)$ is the mean exit time of the Brownian motion $\rvX_t$.

\subsection{Random evolution}

While a deterministic dynamic equation 
coupled to a white noise is called a {\it stochastic differential 
equation} and leads to Brownian motion \cite{Ok}, a deterministic 
dynamic (evolution) equation coupled to a Markov process is called 
{\it random evolution} \cite{Pi}.  This is a class of stochastic 
models characterized by a system of equations like
\begin{eqnarray}
   \frac{\partial P(x,-)}{\partial t} &=& 
     - \frac{\partial}{\partial x}
       \left( F_-(x)P(x,-)\right)
	      - f(x)P(x,-) + g(x)P(x,+) 		\nonumber\\
   \frac{\partial P(x,+)}{\partial t} &=& 
     - \frac{\partial}{\partial x}
       \left( F_+(x)P(x,+)\right)
                 + f(x)P(x,-) - g(x)P(x,+).         \nonumber\\
						\label{ranevo}
\end{eqnarray}    
There is no diffusive motion in the movement.  A particle 
follows deterministic ODEs $\dot{x}=F_{\pm}(x)$ and jumps
between (+) and ($-$) states.  Equations like (\ref{ranevo}) have
wide applications in chemistry and biology.  For example, the 
stochastic averaging problem in nuclear magnetic resonance 
spectroscopy is precisely such a problem \cite{PH,Q6}.  
For a recent work, see \cite{EK}.

	For large $f$ and $g$, the motion is approximately $\dot{x}$ 
= $\frac{g(x)F_-(x)+f(x)F_+(x)}{g(x)+f(x)}$.  For extremely
large $f$ and $g$, the Markovian process approaches a rapidly varying
white noise and (\ref{ranevo}) again approaches a diffusion 
equation \cite{Q6}.  In spectroscopy, this corresponds to 
two distinct spectral lines merging into a single broad peak.  

	For small $f$ and $g$, if both $F_-(x)$ and $F_+(x)$ have zeros, 
then the motion of the particle is still qualitatively simple:  the 
particle will stay at a (+) fixed point, jump to ($-$), relax to a 
($-$) fixed point, and stay there until jumping to (+) and relaxing
back to the (+) fixed point, or relaxing to another (+) fixed point. 
If the two $F$'s are arranged appropriately, the particle can be 
continuously unidirectionally transported, step by step, as demonstrated 
in \cite{AB1}.

	One insight from this discussion is that though Eqn (\ref{ahe2})
has no appropriate stationary solution, the time-dependent solution
should be well behaved.  This points out that one should approach 
the time-dependent version of (\ref{ahe2}) rather than its stationary
solution (Eqn. \ref{ss}).

\subsection{Small diffusion and the theory of large deviations} 

Is the dynamics of the degenerate equation (\ref{ranevo}) the limiting 
behavior of Eqn. (\ref{ahe1}) when $D_{\pm} \rightarrow 0$? 
This is clearly an important mathematical question which also 
has significant relevance to the modeling of muscle contractions.  
As we have stated, one way to address this question is to develop a 
complete singular perturbation theory for (\ref{ahe1}).  There is, 
however, also a stochastic approach called {\it theory of 
large deviations} \cite{EF} which in recent years has thrown 
much light on the problem.  A combination of both approaches 
is undoubtedly desirable.  This technically very demanding task 
has been carried out in several occasions, for example in \cite{MS}. 
For a review, see \cite{Sc}.

\section{Deterministic vs. Stochastic Motion of Molecular Motors}
\noindent
While there is no doubt that the motor protein movement is a 
drifted Brwonian motion, the extent of the randomness in the motion
can be quantitatively characterized according to the mathematical 
model.  Let's now again consider Eqn. (\ref{ahe1}) in which there
is no force in the detached state ($-$).  As we have discussed above, 
in the limit of both $D_{\pm} = 0$, the motion will be trapped 
at the zeros of $F(x)$.  This indicates the importance
of nonzero $D_-$ for the motor movement in this model, as has been
repeatedly pointed out by Peskin et al \cite{POO,PEO}.   

	On the other hand, when $F_-(x)\neq 0$, a motor protein 
can move strictly in one direction in a random evolution model.  
Diffusion plays no role in this mechanism. These two 
different modes of movement correspond nicely with ``Brownian 
ratchet'' and ``power stroke'' in the biochemical literature. 
Whether a motor protein in fact moves back-and-forth with a 
drift or almost unidirectional consecutively can be 
quantitatively analyzed.  Until now, there has no quantitative 
means to differentiate these two types of motion.  The present 
theory offers a quantity method to address this issue.  Taking 
Eqn. (\ref{ahe0}) as an example, we can introduce a function 
\begin{equation}
    \int_0^L \left[\left(D\frac{dP(x)}{dx}\right)^2 
       + \left(\frac{F(x)P(x)}{\beta}\right)^2\right]dx
\end{equation}
as the total movement of the protein.  Note that the first term
is associated with the Brownian motion and the second term is 
associated with the unidirectional movement.   Hence their ratio
quantitatively characterizes the mode of the motor movement.  
This integral is known as action in the theory of large deviations
\cite{FW}.  

\section{Future Work}
\noindent

While the detailed mathematical analyses remain to be carried out 
for models of single motor movement, the mathematic analysis of a 
chain of motor protein is largely unknown, except for the completely 
rigid chain of motors (Huxley model).  The general theory can be developed 
by connecting $N$ motor proteins by springs.  Such a ``bead-and-spring''
model has been the theoretical foundation of polymer physics
\cite{DE,Q8}. The late Professor P.J. Flory was awarded the Nobel Prize in 
chemistry in 1974 for his contribution to this theory.  The
difference, however, is that a polymer is an equilibrium system, while a
chain of motors is a ``living creature''.
Let's denote the positions of $N$ motor proteins by $x_1$, $x_2$, ...,
$x_N$, and the corresponding internal states by $\sigma_1$,
$\sigma_2$, ..., $\sigma_N$, where $\sigma_k =0, 1$ for the detached and
attached state of the $k$th motor.  We therefore have the dynamic
equation for probability
$P(x_1, \sigma_1, x_2, \sigma_2,...,x_N,\sigma_N,t)$:

\[ \frac{\partial P}{\partial t} =
   \sum_{k=1}^N \left[  \{(1-\sigma_k)D_- + \sigma_k D_+\}
   \frac{\partial^2 P}{\partial x_k^2}
       - \frac{\partial}{\partial x_k} \left(
         \frac{\sigma_k F_-(x_k)}{\beta_-} +
         \frac{(1-\sigma_k)F_+(x_k)}{\beta_+} \right. \right.  \]
\[  +\left. \left\{ \frac{1-\sigma_k}{\beta_+} +
   \frac{\sigma_k}{\beta_-} \right\} \eta(x_{k-1}-2x_k+x_{k+1}) 
	P\right) 		  \]
\[ -\{(1-\sigma_k)f(x)+\sigma_k g(x)\}P(...,\sigma_k,...)   \]
\begin{equation}
     +\{(1-\sigma_k) g(x)+\sigma_k)f(x)\}P(...,1-\sigma_k,...) \Bigg].
\label{nmotor}
\end{equation}
where $\eta$ is a spring constant.  A computational analysis of
such a model can be found in \cite{DTC}.  In a recent mathematical
analysis, a deterministic counterpart of this system, a chain 
of bead-and-spring in a periodic force field, has been shown  
to exhibit globally phase-locked motion \cite{QZQ,QW2}.  
Eqn. (\ref{nmotor}) is a N-particle system which can be 
subjected to mean-field treatment as that for the N-particle
Sch\"{o}rdinger equation.  As in the genesis of nonlinear Schr\"{o}dinger 
equation \cite{Rein}, such treatment will lead to a nonlinear 
diffusion equation \cite{GPV}, opening a possible new 
mathematical approach to the problem of muscle contraction.  In 
connection to the theory of probability, this is an interacting 
particle system with a nonequilibrium (Gibbsian) stationary 
state \cite{Li}, and is a natural application for the theory 
of large deviations \cite{DZ}.

\vskip 0.5cm\noindent
{\bf\Large Acknowledgements}
\vskip 0.3cm

I like to thank numerous colleagues in biochemistry for many helpful 
discussions.  I am grateful to my mathematical colleagues, Professors 
Zhen-qing Chen, Randy LeVeque, Jim Murray, and Bob O'Malley for 
comments on the manuscript.


\end{document}